\documentclass{emulateapj}
\usepackage{aas_macros}
\usepackage{epstopdf}
\usepackage{epsfig}
\usepackage{natbib}
\usepackage{float}
\usepackage{gensymb}
\usepackage{amsmath}
\usepackage{lipsum}% http://ctan.org/pkg/lipsum
\usepackage{times}
\usepackage{comment}
\usepackage{color}
\usepackage{multirow}
\usepackage{soul}
\usepackage{CJKutf8}            % Chinese name

\definecolor{purple}{RGB}{160,32,240}

\definecolor{purple2}{RGB}{120,72,240}

\newcommand{\rmag}{\>^{0.1}{\rm M}_r-5\log h}

\begin{document}

%\title{Emission of the Ionized Gas in Group of Galaxies}
\title{H$\alpha$ Emission and the Dependence of the Circumgalactic Cool Gas Fraction on Halo Mass}

\author{Huanian Zhang \begin{CJK*}{UTF8}{gkai}(张华年)\altaffilmark{1}, Xiaohu Yang  (杨小虎)\end{CJK*}\altaffilmark{2,3}, Dennis Zaritsky\altaffilmark{1}, Peter Behroozi \altaffilmark{1}, and Jessica Werk \altaffilmark{4}}
\altaffiltext{1}{Steward Observatory, University of Arizona, Tucson, AZ 85719, USA; fantasyzhn@email.arizona.edu}
\altaffiltext{2}{Department of Astronomy, School of Physics and Astronomy, Shanghai JiaoTong University, Shanghai, 200240, China}
\altaffiltext{3}{Tsung-Dao Lee Institute, and Shanghai Key Laboratory for Particle Physics and Cosmology, Shanghai Jiao Tong University 
  Shanghai, 200240, China}
  \altaffiltext{4}{Department of Astronomy, University of Washington, Seattle, WA 98195,  USA}

\begin{abstract}
We continue our empirical study of the emission line flux originating in the cool ($T\sim10^4$ K) gas that populates the halos of galaxies and their environments.  Specifically, we present results obtained for a sample of nearly half a million individual galaxies, groups, and clusters of galaxies, intersected by more than two million SDSS lines of sight at projected separations of up to a quarter of the virial radius. 
Adopting simple power law relationships between the circumgalactic (CGM) cool gas fraction and either the halo or stellar mass, we present expressions for the CGM cool gas fraction as a function of either halo or stellar mass, %$C_f(M_h) = (0.23^{+0.07}_{-0.06}) \times (M_h/10^{12}M_\odot)^{(-0.41^{+0.07}_{-0.09})} + (0.028^{+0.030}_{-0.020})$ 
$f_{\rm cool}(M_h) = (0.25^{+0.07}_{-0.06}) \times (M_h/10^{12}M_\odot)^{(-0.39^{+0.06}_{-0.07})}$
or %$C_f(M_{*}) = (0.25\pm 0.07) \times (M_{\rm *}/10^{10.0}M_\odot)^{(-0.35\pm 0.07)} + (0.026^{+0.032}_{-0.019})$. 
$f_{\rm cool}(M_{*}) = (0.28^{+0.07}_{-0.06}) \times (M_{\rm *}/10^{10.0}M_\odot)^{(-0.33\pm 0.06)}$.
Where we can compare, our results are consistent with previous constraints from absorption line studies, our own previous emission line work, and simulations.  
%We also set constrains on the gas temperature to be 9257$^{+7295}_{-4071}$ K.
The cool gas can be the dominant baryonic CGM component, comprising a fraction as high as $> 90\%$ of halo gaseous baryons, in low mass halos, $M_h\sim$ $10^{10.5} M_\odot$, and a minor fraction, $<$ 5\%, in groups and clusters, $M_h > 10^{14} M_\odot$.

\end{abstract}


\section{Introduction}

Galaxies are surrounded by extended and diffuse gas, referred to as the circumgalactic medium (CGM), which is a critical but incompletely understood part of galactic ecosystems \citep{CGM2017}. Previous studies \cite[e.g.,][]{Behroozi2010,McGaugh2010} concluded that only $\sim$ 20\% of all the baryons apportioned to a given halo have been converted into stars for $L^*$ galaxies. The corresponding fractions are even lower for sub-$L^*$ and super-$L^*$ galaxies, typically only $\sim$ 5$-$10\%. The common inference is that the majority of the original baryons in a given halo remain in the halo and comprise the CGM.

The CGM is a multi-phase medium with rich dynamics and complex ionization states \citep{Bregman,Werk2014}. It  provides the fuel for subsequent star formation activities \citep{spitzer} and  serves as the depository for galactic recycling and feedback \citep{CGM2017}. Empirical constraints on the nature of the CGM have come primarily from the study of absorption lines in the specrta of bright background objects \citep[e.g.][]{steidel2010,menard2011,bordoloi2011,zhu2013a,zhu2013b,Werk2014,werk16,croft2016,croft2018,prochaska2017,Cai2017,lan2018,joshi2018}. Unfortunately, absorption line studies are limited by the requirement of a sufficiently bright background source and so typically probe a single sight line through each of a limited set of halos.
As such, a valuable complement to these studies is the detection and study of emission line flux from the CGM, which will eventually provide spatially resolved measurements of the CGM of individual galaxies and, by doing so, help resolve modeling degeneracies and provide novel constraints.  

\cite{zhang2016} presented the first detection of the emission line
flux, H$\alpha$+N{\small[II]}, from low redshift, normal galaxies extending out to $\sim$ 100 kpc projected radius. This measurement could only be achieved with existing data by stacking a sample of over 7 million lines of sight from the Sloan Digital Sky Survey \citep[SDSS DR12;][] {SDSS12}. 
Building on that first result, we have applied the technique to characterize the nature of the emission and the CGM in low redshift galaxies and their environments \citep[][hereafter, Papers I, II, III, and IV]{zhang2016,zhang2018a,Zhang2018b, Zhang2019}.

In those studies, our focus was on the nature of line emission from the halos of ``normal" galaxies. As such, we purposefully limited
the range of galaxy masses and sizes considered. 
Now, we expand the range of halo masses considered as much as possible to characterize the cool gas halo component as a function of halo mass. The virial temperature of the halo gas can be as low as $10^{4-5}$ K for small halos with mass  $\lesssim 10^{11} M_\odot$ and as high as $10^{6-7}$ K
for large halos with  mass $\gtrsim 10^{13} M_\odot$.
Given this large range of virial temperatures, one naturally expects a strong trend in the fraction of the CGM that is in the  cool gas phase (T $\sim 10^4$ K) as a function of mass.

Our aim here is to 
build on existing evidence from absorption line studies and detailed theoretical modeling for such a trend \citep{Werk2014,Ford2014,suresh2017} 
%and to 
%further
%test this hypothesis 
and refine the quantitative description of the trend
using stacked emission line measurements.
Throughout the paper we adopt a $\Lambda$CDM cosmology with parameters
$\Omega_m$ = 0.3, $\Omega_\Lambda =$ 0.7, $\Omega_k$ = 0 and the dimensionless Hubble constant $h = $ 0.7 \citep[cf.][]{riess,Planck2018}.

\section{Data Analysis}

\subsection{The Parent Sample}

We begin by describing the set of systems about which we will measure the emission flux from the surrounding cool gas. Because we are expanding the halo mass range explored relative to our previous studies, our systems now span classification from individual galaxies, to galaxy groups, and finally to galaxy clusters. For simplicity, we will refer to all systems with multiple luminous galaxies as galaxy groups, even those that would generally be referred to as galaxy clusters. In this regard, we utilize an existing ``group" catalog \citep[hereafter Y12]{Yang2012}\footnote{see {\tt http://gax.sjtu.edu.cn/data/Group.html}}, which is based on the work of \cite{yang2007}.

We briefly describe how that catalog was constructed for completeness. Y12 selected galaxies with
redshifts in the range $0.01 \leq z \leq 0.20$, and a redshift
completeness\footnote{Because of the fiber-collision effects, some neighboring galaxies in the SDSS observations do not have spectroscopic redshifts. This effect becomes increasingly important in dense environments, so we set the redshift completeness criteria.}  ${\cal C}_z > 0.7$ from the main galaxy sample of the New York
University Value-Added Galaxy Catalog \citep[NYU-VAGC][]{Blanton2005} for the SDSS DR7 \citep{Abazajian2009}.
This selection results in a sample of $639,359$
galaxies scattered over 7748 square degrees. From this galaxy sample, a galaxy group catalog was constructed using an adaptive halo-based group finder \citep[see][for details]{yang2007, Yang2012}. 
For each identified system, whether it is a single galaxy or a multiple galaxy system, the dark matter halo
mass, $M_h$, is estimated using the rank of the system's total stellar mass, $M_{\rm stellar}$, which is defined to be the sum of the stellar mass of all members with $\rmag \leq
-19.5$, and a ranked halo mass function. {\bf Meanwhile, we take into account the redshift completeness limit of the groups, and adopt the halo mass function for a given cosmology  to determine the halo mass of the most massive halo expected within this volume, the second most massive, and so on. We then assign the group with the largest stellar mass the most massive halo and work our way down the rankings. The details of such a halo mass estimation algorithm is described in detail in \S3.5 of Y07.} As Y07 show, the estimated  masses 
recover the true halo masses with a 1-$\sigma$ deviation of $\sim 0.3$ dex for $L^*$ galaxies, although the scatter at the small mass end is unknown \citep{Allen2018}, 
and are more reliable than those based on the velocity dispersion of group
members \citep{Yang2005, Weinmann2006,  Berlind2006}. 
For those groups where all the member galaxies  are fainter than $\rmag =
-19.5$, the halo masses are estimated according to the stellar to halo mass relation for central galaxies obtained in Y12. 

The resulting catalog, constructed from 639,324 galaxies, contains 472,416 identified systems, of which 23,735 are groups with three or more galaxy members, 44,470 are binaries, and the remaining 404,211 are isolated galaxies. %In summary, there are 146,173 galaxies that are members of $N\ge 3$ groups, 88,940 galaxies in binary systems, and the remaining 404,211 galaxies are classified as isolated.
%\Jess{Jess: I think it's nice to give the precise numbers here, but when they don't obviously add up to the "total number of systems" it seems a bit odd to present this way. Okay, so: 23,735 + 88,940/2 + 404,211 =  472416 -- maybe just add to the top, 23,735 are groups with 3 or more, 44,470 are binary systems, and the remaining 404,211 are isolated (that adds up nicely, obviously in terms of total "systems").  Then you can give the actual TOTAL number of galaxies independently ($>$ 600,000). } \Huanian{resived}
A critical property of the catalog is that systems with only one member are not satellite galaxies associated with larger galaxies or groups, they are independent, parent halos. The isolated galaxies may have satellites that are fainter than the spectroscopic limit or companions that may have been missed due to spectroscopic incompleteness above the magnitude limit. The mean and median derived halo masses for $N=3$ groups are $10^{12.86}$ and $10^{13.04}$ $M_\odot$, respectively.  About 10\% (42,198) of the isolated galaxies are assigned halo masses greater than that of the mean $N=3$ group. 
For the halo mass range where individual and multi-galaxy systems overlap, the resulting H$\alpha$ fluxes that we measure as described below are within the 1$\sigma$ uncertainties whether we include or exclude the massive individual galaxies. We detect no difference in the CGM emission properties of isolated and group galaxies in this overlapping mass range and provide quantiative support for this claim in \S3. %\Jess{Jess: this seems like a crucial point, and it just sort of sits here without any reference or figure. Can you expand on this, or back it up? It is very interesting...} \Huanian{The samples without those isolated massives give the measurements for the two mass bins as following: ($0.0024\pm0.0012$, $0.0012\pm0.0005$), ($0.00085\pm0.0014$, $-0.0002\pm0.0006$), almost identical to the values in Table 1. } 

\begin{figure}[htbp]
\begin{center}
\includegraphics[width = 0.48 \textwidth]{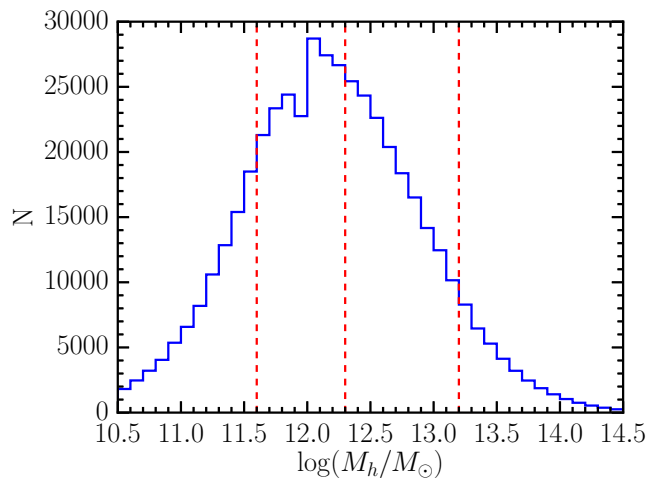}
\end{center}
\caption{The distribution of halo masses in the catalog. The dashed vertical lines show the boundaries we set between the four mass bins we use in our analysis.} 
\label{fig:dist_halom}
\end{figure}

The halo masses for the cataloged objects range 
from $10^{10.7} M_\odot$, for systems with only one member galaxy, to $10^{15.0} M_\odot$, for systems with $>$ 500 member galaxies.  We display the distribution of the halo masses for the cataloged systems in Figure \ref{fig:dist_halom}. The dashed red lines indicates the boundaries of four different mass bins that we define. We select these bins to correspond broadly to halos similar in mass to that of 1) the Large Magellanic Cloud \citep[$\sim 10^{11.40} M_\odot$;][]{LMC2016},  
2) the Milky Way \citep[$\sim 10^{12}$ M$_\odot$;][]{z89}, 3) massive galaxies like M 87 \citep[$\sim 10^{13}$ M$_\odot$;][]{m87},
and 4) everything else from poor groups to the most massive clusters
\citep[$\gtrsim 0.5 \times 10^{14}$ M$_\odot$;][]{zab}.
At the same time, we are trying to balance the number of halos in each  mass bin as much as possible. The average halo mass of objects in each of the four mass bins are $10^{11.27}$, $10^{11.97}$, $10^{12.72}$ and $10^{13.75} ~ M_\odot$. 

\subsection{The Stacking Analysis}

In expanding our analysis from a limited sample of galaxies in Papers I-IV to the wider variety of systems considered now, we introduce several complications. First, while the center for an individual galaxy is fairly straightforward to identify, the center of a group or cluster is more ambiguous. Second, with a sample of fairly similar galaxies, we could stack our results in terms of physical projected radius. For our current sample, we cannot stack on physical projected radius because a fixed radius probes quite different physical regimes in different halos. Third, in this work we will be measuring composite emission line profiles. For individual galaxies, we probe gas in their halos when we restrict ourselves to $<$ 50 kpc projected separation \citep{zhang2016}, but for groups we are measuring the combination of the cool gas from the parent and satellite halos. As such, our measurements are of the total cool gas in these halos, whether it is physically associated with the parent or satellites. Our previous work \citep{Zhang2019} suggests that emission from satellite halos in dense environments is suppressed, so our measurement may indeed primarily  reflect the gas associated with the parent, but this interpretation cannot be established within this analysis.

In general, we follow the  approach developed in Papers I-IV. 
We obtain galaxy spectra from the Sloan Digital Sky Survey Data Releases \cite[SDSS DR12]{SDSS12} projected within 1 Mpc of the targets described above \citep{Yang2012}. The outer 1 Mpc separation limit is sufficiently large to satisfy the rescaling that we describe further below. 
To avoid contamination from emission arising in the luminous centers of satellites or other nearby galaxies, we exclude SDSS lines of sight that targeted galaxies whose redshift are within 0.05 of our defined target. For each remaining spectrum, we fit and subtract a 10th order polynomial to a 300 \AA\ wide section at the rest wavelength, in the target reference frame, of H$\alpha$ to remove the continuum. {\bf Various tests of the procedure, including the injection and accurate recovery of emission lines of known flux, are described in detail in \cite{zhang2016}.}
Unlike in previous papers, where we measured both H$\alpha$ and [N{\small II}], here we focus solely on H$\alpha$ to simplify the modeling and interpretation.

\subsubsection{Centering}

Defining the sample of spectra and stacking in terms of projected separation requires a definition of the ``center" of each target. For isolated galaxies, that is unambiguous, but for multiple galaxy systems there are at least two options. One can define the center to correspond to the position of the brightest group galaxy (BGG) or to the luminosity weighted center (LWC) of all the identified member galaxies. We will explore both approaches, but  focus on the results using the LWC, which results in centers that are consistently defined across systems that do and do not have a clearly dominant galaxy.

We find that the choice of center is not critical in our analysis. 
The center as defined in these two ways tends to differ at a level of between 1 and 10 kpc for systems with 2 or more members. Given that we do not use the inner 10 kpc of the profile in our analysis, to avoid contamination from emission from the central galaxy, the resulting differences in profiles are expected to be relatively minor. We do, nevertheless, compare our results using both methods even though we present the quantitative results from the LWC approach in figures and 
modeling results. We find that the results we present are insensitive to the choice of centering.

\subsubsection{Rescaling}

Comparing the properties of a set of systems with highly disparate masses and sizes requires some care. Specifically, we need to compare measurements at similar, representative projected radii and over appropriate velocity windows. We do not favor comparing results at physical radii. For example, a projected radius of 25 kpc corresponds to a halo position for a sub-$L^*$ galaxy but is likely to be within the BGG of our most massive clusters. Instead, we will compare results at scaled radii. Likewise, we need to take care how we integrate along the line of sight. Integrating over a velocity window of width 400 km sec$^{-1}$ will include all of the bound gas in a low mass galaxy but only a limited fraction of the gas in a massive cluster.

We choose to measure the emission line fluxes in bins of scaled projected radius ($r_s$), where we scale the physical projected radius ($r_p$) in terms of the virial radius ($r_s \equiv r_p/R_\mathrm{180}$).  We estimate the virial radius ($\equiv R_{180}$) 
%and concentration 
using the following equation from \cite{yang2007}:
\begin{equation}
R_{180} = 780~ h^{-1} {\rm kpc}\Big(\frac{M_h}{\Omega_m 10^{14}h^{-1} M_\odot}\Big)^{1/3} (1+z)^{-1}
\end{equation}
%and
%\begin{equation}
%c = 10.931 - 2.274 x_m + 0.04 x_m^2,
%\end{equation}
where $R_{180}$ is the radius within which the mean halo density is 180 times the current critical density. %$c$ is the concentration of the dark matter halo as introduced in the NFW profile \citep{NFW1996,NFW1997} that we will use later in our modeling, and $x_m \equiv \log M_h/(h^{-1} M_\odot) - 12$ \citep{Maccio2007}.
We will measure the flux for $0.05 \le r_s \le 0.25$, which corresponds to a range of 10 to 50 kpc for an $\sim$ $L^*$ galaxy and matches our previous work using fixed physical radii. For M$_h > 10^{14} M_\odot$, the virial radius is $\gtrsim$ 2 Mpc and then the $r_s$ range corresponds to physical projected radii of 100 to 500 kpc. 
%\Jess{just to note, COS-Halos, COS-Dwarfs, CASBAH, and other impactful absorption line surveys all use R\_{200}. It's not hugely different, but for ease of comparison to those datasets it might be nice to change this if it's not a huge amount of work... ?}
%\Dennis{not critical to change here - none of our measurements are out to the virial radius - we just use it to scale - and since our scaling is built from physical radii (i.e. trying to stay between 10 and 50 kpc for L* galaxies, we would simply use a slightly different multiplier relative to R$_{200}$ and it still wouldn't be directly comparable to quoted numbers in those other studies.}
 
We measure the emission flux within a prescribed velocity window relative to the target in each  individual spectra that probes the halo and then combine these measurements.
Because $M_h$ spans a few orders of magnitude, we need to vary the size of the velocity window. For targets in each of the four mass bins, we scale the window roughly to the characteristic virial velocity of targets in the bin, adopting windows of $\pm$ 100 km s$^{-1}$,   $\pm$ 215 km s$^{-1}$, $\pm$ 330 km s$^{-1}$, and $\pm$ 450 km s$^{-1}$. At the ends of the virial velocity distribution we have too large a window for our least massive objects and too small a window for our most massive due to the long tails of the mass distribution (Figure \ref{fig:dist_halom}).

This limiting of the velocity window scaling at low and high velocities does not affect our results. At the low velocity end our window is larger than that we would have if we continued our scaling downward. Only in highly crowded environments might an overly large velocity window allow flux from neighbors to contaminate the measurement. However, these targets are almost certainly isolated, since any neighbor is likely to be more massive and this target would then not have been categorized as isolated. To confirm that this windowing of the lowest mass galaxies is not affecting our results, we redo our measurement of the H$\alpha$ flux in the lowest mass bin excluding all systems with $M_{\rm halo} < 10^{11} M_\odot$.  The measured flux drops from $0.011 \pm 0.004$ to $0.0082 \pm 0.0044$, but the new value is statistically consistent with the original one. This test demonstrates that the effect of a large window on these lowest mass systems is not resulting in a systematic error that is larger than the statistical errors.  
At the high end of halo mass range, our window is not as large as it should be and we could be losing flux. We had to limit the size of the window to avoid contaminating the H$\alpha$ measurement with [N {\small II}] emission. {\bf This criteria limits the maximum velocity difference between the central galaxy and H$\alpha$ CGM emission to $<$ 450 km sec$^{-1}$, which is below the virial velocity for the more massive systems.}  To test for the choice of window size on the measured H$\alpha$ fluxes in the largest M$_h$ bin, we now measure the flux in a velocity window where the blue end of the spectral window is defined by the full characteristic virial velocity and at the upper by 0 relative velocity. We would expect such a choice of velocity window to result in a measurement of half of the total flux.
We can do this as a test because the [N {\small II}] line that is blueward of H$\alpha$ is much weaker than the redder one and is rarely detected in our stacks \citep{zhang2016}. {\bf By then doubling the flux in this blueward-extended velocity window (our lower velocity limit is set to $-$1350 km s$^{-1}$), we have an estimate of what we would measure across the full window if we did not have to limit its extent in either direction.} 
We find that these new measurement {\bf is $-0.00030 \pm 0.0013$, to be compared to the original measurement of $0.00080 \pm 0.0011$ for the smaller $r_s$ bin, and is $-0.00006 \pm 0.00066$, to be compared to $0.00039 \pm 0.00037$, for the larger $r_s$ bin. The test results} are statistically consistent with our previous measurements. We conclude that although not ideal, our truncated velocity window is not introducing systematic uncertainties that are larger than our internal statistical uncertainties in our largest $M_h$ bins. 

We also appeal for justification of our choice to absorption line studies that find that the CGM gas is typically found at velocities well below the escape speed. \cite{Werk2013} and \cite{Tumlinson2013} find that most of the absorption is within 150 km sec$^{-1}$ of the galaxy's systemic velocity even when the escape speed was estimated to be 300 km s$^{-1}$. Nevertheless, one might expect some gas to be escaping, particularly in the lower mass systems \citep[e.g.,][]{Bordoloi2014}. \cite{Tripp2011} find a few examples of absorption line systems offset in velocity by more than the escape speed. This gas could be interpreted as escaping or perhaps it is associated with other galaxies, as we find gas at large projected radii to be \cite{zhang2018a}. Even if it is escaping gas, it is not evident that we would want to be including it in our measurement if our aim is to quantify the gas reservoir for future star formation. 

\section{Results}

We present stacked measurements of the H$\alpha$ fluxes in two $r_s$ bins (the inner bin spans $0.05 < r_s < 0.11$ and the outer bin spans $0.11 < r_s < 0.25$) as a function of $M_h$ and $M_*$ in Table \ref{tab:Rdata}\footnote{
The conversion factor to units between the values we present,  $10^{-17}$ erg cm$^{-2}$ s$^{-1}$ \AA$^{-1}$ and those used commonly in the literature to describe diffuse line emission, erg cm$^{-2}$ s$^{-1}$ arcsec$^{-2}$, is 1.7.}. We do not extend the radial range beyond $r_s = 0.25$ because the H$\alpha$ flux beyond $r_s = 0.25$ for $\sim L^*$ galaxies becomes increasingly dominated by the contribution from associated halos, not the target galaxy itself \citep{zhang2018a}.
In figures and tables we present the median emission line fluxes and the associated  uncertainties, estimated using a jackknife method.  Specifically, we randomly select half of the individual spectra, calculate the mean emission line flux and repeat the process 1000 times to establish the distribution of measurements from which we quote the values corresponding to the 16.5 and 83.5 percentiles as the lower and upper uncertainties, respectively. Returning to the question of whether the emission flux is similar for similarly massive single or multiple galaxy systems that we first discussed in \S2.1, we note that if we remove the isolated galaxies from the bin with $\langle M_h \rangle = 10^{12.72}$ M$_\odot$, the new values of the fluxes in the inner and outer radius bins are $0.0024 \pm 0.012$ and $0.0012 \pm 0.0005$, respectively, and nearly indistinguishable from the values in Table 1.

\begin{deluxetable*}{ccccc}
\tablewidth{0pt}
\tablecaption{The emission fluxes for H$\alpha$ as a function of radius and halo mass/stellar mass$^a$}
\tablehead{
\\
\colhead{$r_s$} & \multicolumn{4}{c}{Halo Mass}\\ }
\startdata
\\
  &$10^{11.27} M_\odot$ & $10^{11.97} M_\odot$ & $10^{12.72} M_\odot$ & $10^{13.75} M_\odot$ \\
 \\
 \hline \\
0.08 &  $0.011 \pm 0.004$ & $0.0065\pm 0.0018$ & $0.0025\pm 0.0012$ & $-0.0003\pm 0.0013$\\
\\
 0.2 & $0.0055 \pm 0.0015$ & $0.0019 \pm 0.0007$ & $0.0010 \pm 0.0005$ & $-5\times 10^{-5} \pm 0.0006$ \\
 \\
 \hline \\
 &\multicolumn{4}{c}{Stellar Mass}\\
 \\
 \hline\\ 
 &$10^{9.33} M_\odot$ & $10^{10.18} M_\odot$ & $10^{10.84} M_\odot$ & $10^{11.60} M_\odot$ \\
 \\
 \hline\\
 0.08 &  $0.0057 \pm 0.0046$ & $0.010\pm 0.002$ & $0.0022\pm 0.0010$ & $0.0002\pm 0.0014$\\
\\
 0.2 & $0.0059 \pm 0.0021$ & $0.0023 \pm 0.0008$ & $0.00087 \pm 0.00050$ & $-0.0001 \pm 0.0006$ 
\enddata
\label{tab:Rdata}
\tablenotetext{a}{Fluxes in units of $10^{-17}$ erg cm$^{-2}$ s$^{-1}$ \AA$^{-1}$}
\end{deluxetable*}

\subsection{Emission Fluxes as a Function of System Mass}

\begin{figure}[htbp]
\begin{center}
\includegraphics[width = 0.48 \textwidth]{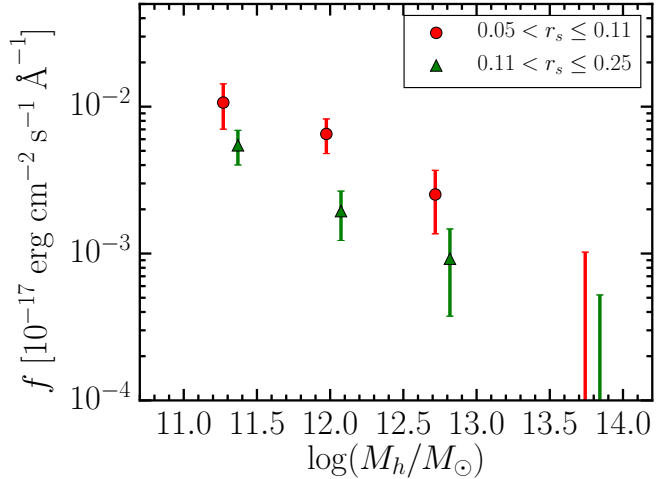}
\end{center}
\caption{The H$\alpha$ emission fluxes at two values of $r_s$ for four different halo mass ranges. For visualization, we apply slight horizontal offsets to the two samples although they have the same halo mass.}
\label{fig:halom}
\end{figure}

\begin{figure}[htbp]
\begin{center}
\includegraphics[width = 0.48 \textwidth]{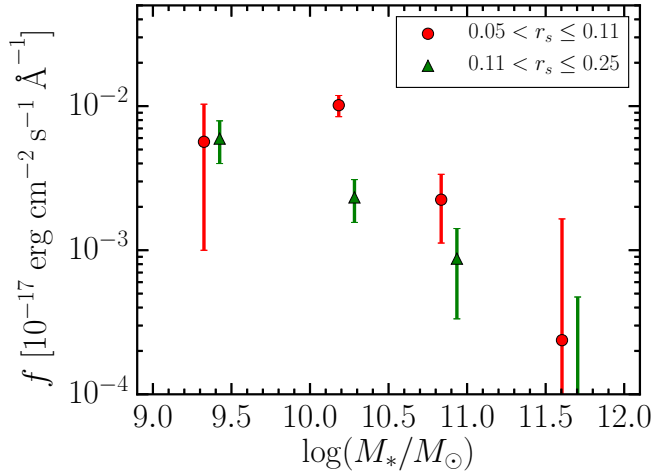}
\end{center}
\caption{The H$\alpha$ emission fluxes at two values of $r_s$ for four different stellar mass ranges. For visualization, we apply slight horizontal offsets to the two samples although they have the same stellar mass.}
\label{fig:sm}
\end{figure}

In Figure \ref{fig:halom} we present the H$\alpha$ emission fluxes at two different scaled radii as a function of halo mass {\bf and in Figure \ref{fig:sm} as a function of stellar mass}. We see a systemic decline in the flux at both radii as the masses increase. {\bf Even had the flux remained flat as a function of mass, the fraction of the baryonic reservoir that this represent would be declining rapidly because the halo mass is increasing by a factor of $\sim$ 300 across the plotted mass ranges.}
These trends suggest a strong inverse relationship between the fraction of the CGM in the cool component vs. halo mass. 

%When we split the data for different  stellar mass bins (Figure \ref{fig:sm}), we obtain qualitatively consistent results with those for halo mass. 

\subsection{Modeling the Cool Gas Fraction vs Halo Mass}

The results in Figure \ref{fig:halom} imply that the cool gas fraction, $C_f$, declines strongly with $M_h$, but they do not directly tell us what that dependence is. To place constraints on that relation we turn to a straightforward modeling tool that we employed previously in Papers II and IV.  
The methodology is based on the UniverseMachine architecture \citep{Behroozi2018} where standard $\Lambda$CDM dark matter halos within a cosmological volume are populated with luminous galaxies in a manner to match available observational constraints. We then apply a hypothesized behavior for the dependence of the cool CGM gas fraction on halo mass and produce mock observations where the selection function is matched closely to the real observations. The parameters of that hypothesized behavior are constrained by producing a model that can reproduce the observations.

Some assumptions are needed to populate the modeled halos with the gas that produces the H$\alpha$ emission. First,  
we adopt an analytic description for the gas density profile within a dark matter halo derived for gas in an NFW potential \citep{NFW1996,NFW1997} in  hydrostatic equilibrium \citep{Capelo2010}. 
Second, we need to adopt a certain range of temperatures for the cool gas component that we are modeling. In previous papers, we drew on a rough estimation of the temperature as $10^4$ K. This choice is supported by studies finding that the temperature distribution is bimodal, with one peak around $\sim 10^6$ K and another centered at $\sim 10^4$ K \citep{Haider2016,Cui2019}. 
The cool gas temperature, $T$, is an important parameter as it sets the recombination rate, $\beta_{\rm H\alpha}$,
\begin{equation}
\beta_{\rm H\alpha} = 10^{-13} \frac{2.274 ~T^{-0.659}}{1+1.939 ~T^{0.574}}
\label{eq:recombination}
%\end{split}
\end{equation}
\citep{Pequignot1991}.
While the assumption of a single characteristic temperature, with some allowed scatter, is defensible when modeling a narrow range of systems, it becomes more questionable as one explores the large range of systems we are dealing with here. We will discuss this issue further below. 

Once the shape of the radial density profile and temperature of the cool gas are defined, we need to specify the fraction of CGM that is in this phase.
Our cool gas fraction, $f_{\rm cool}$\footnote{Here we denote cool gas fraction as $f_{\rm cool}$ instead of $C_f$, because $C_f$ in CGM absorption measurements refers to recovering fraction.}, refers, as in our previous papers, to the fraction of CGM baryons that are in the $T\sim10^4$ K phase. We calculate this fraction starting
with the assumption that each halo contains its fair share of baryons, the cosmological baryon fraction or 16\% \citep{Planck2018}, that 80\% of those baryons are in the form of CGM \citep[as inferred from the deficit between the identified stellar plus gaseous mass and halo mass;][]{Behroozi2010}, and that 75\% of those, by mass, are hydrogen (ionized or neutral, {\bf adopting the standard universal hydrogen mass fraction}). The cool gas fraction is then the percentage of the hydrogen that must be at $\sim 10^4$ K to generate the observed H$\alpha$ flux. 
We model the behavior of  $f_{\rm cool}$ with halo mass as a power law, 

\begin{equation}
%\begin{split}
f_{\rm cool}(M_h)  = a \times \Big(\frac{M_h}{10^{12}M_\odot}\Big)^b
\label{eq:halo mass}
\end{equation}

\noindent
where $M_h$ is the halo mass, ($a$, $b$) are fitting parameters. We adopt a scatter of 0.3 dex to the halo mass, {\bf which is consistent with Y07 although the scatter at the small mass end is unknown \citep{Allen2018},} to simulate the halo mass uncertainty resulting from the conversion of stellar mass to halo mass. To be specific, we estimate the cool gas fraction using Eq. \ref{eq:halo mass} with the true halo mass, and then we add scatter to the halo mass when placing each system into its corresponding mass bins.

We perform a Bayesian analysis to derive confidence intervals on each of the model parameters, $(a,b,T)$.
The prior on $a$ we take from our previous work, where the $f_{\rm cool} \sim 0.3$ for $\sim L_*$ galaxies, so we allow $0 < a < 0.5$. 
The prior on $b$ is informed by the 
previous determination, based on absorption line observations, that the cool gas fraction decreases as the galaxy mass increases \citep{Ford2014}. So we adopt 
that the slope $b$ must be negative. 
Finally, we adopt a large range of allowed temperatures. In all cases, we adopt uniform priors within the allowed ranges. To summarize, our priors are:
\begin{equation}
\begin{split}
0 < a < 0.5, \quad -1 < b < 0,   \\   3000 < T < 30000, \quad a\times 0.025^b < 1
\end{split}
\label{eq:prior}
\end{equation}
The additional condition that we apply, $a\times 0.025^b < 1$, is imposed so that the cool gas fraction of any halo is less than one. The posterior distribution, $p(\Theta|{\rm data})$, for the parameters based on the data is described as follows:
\begin{equation}
p(\Theta|{\rm data}) = \frac{p(\Theta) \cdot \Pi ~p({\rm data}|\Theta)}{p({\rm data})}
\end{equation}
where $\Theta$ is the parameter space $\Theta = (a, b, T)$, $p(\Theta)$ is the prior distribution described in Eq. \ref{eq:prior},  $p({\rm data}|\Theta)$ is the likelihood of the data, and $p(\rm data)$ is the marginal probability for the data. 

We define the likelihood of obtaining the data given a specific model using the difference between the actual and model data, ${\rm exp[-0.5\times (actual - model)^2/\sigma^2}]$, where $\sigma$ is the observational uncertainty. We fit the data for the four mass bins at the two values of $r_s$. We use the software package called ``{\it emcee}" \cite[]{emcee}, which implements a Markov chain Monte Carlo sampling of the likelihoods across parameter space to calculate the posterior distribution. $emcee$ is a $Python$ implementation of the affine-invariant ensemble sampling approach suggested by \cite{Goodman2010MCMC}. It utilizes an ensemble of $N ~walkers$, and it will evolve each for a certain number of steps. We initialize each walker (total 500 walkers in our simulation) by randomly sampling from our prior distributions, and evolving each walker for 1500 steps. We discard the first 250 steps since it takes a certain number of steps ($\sim 100$ in this study) for the results to become stable. 

\begin{figure*}[htbp]
\begin{center}
\includegraphics[width = 0.8 \textwidth]{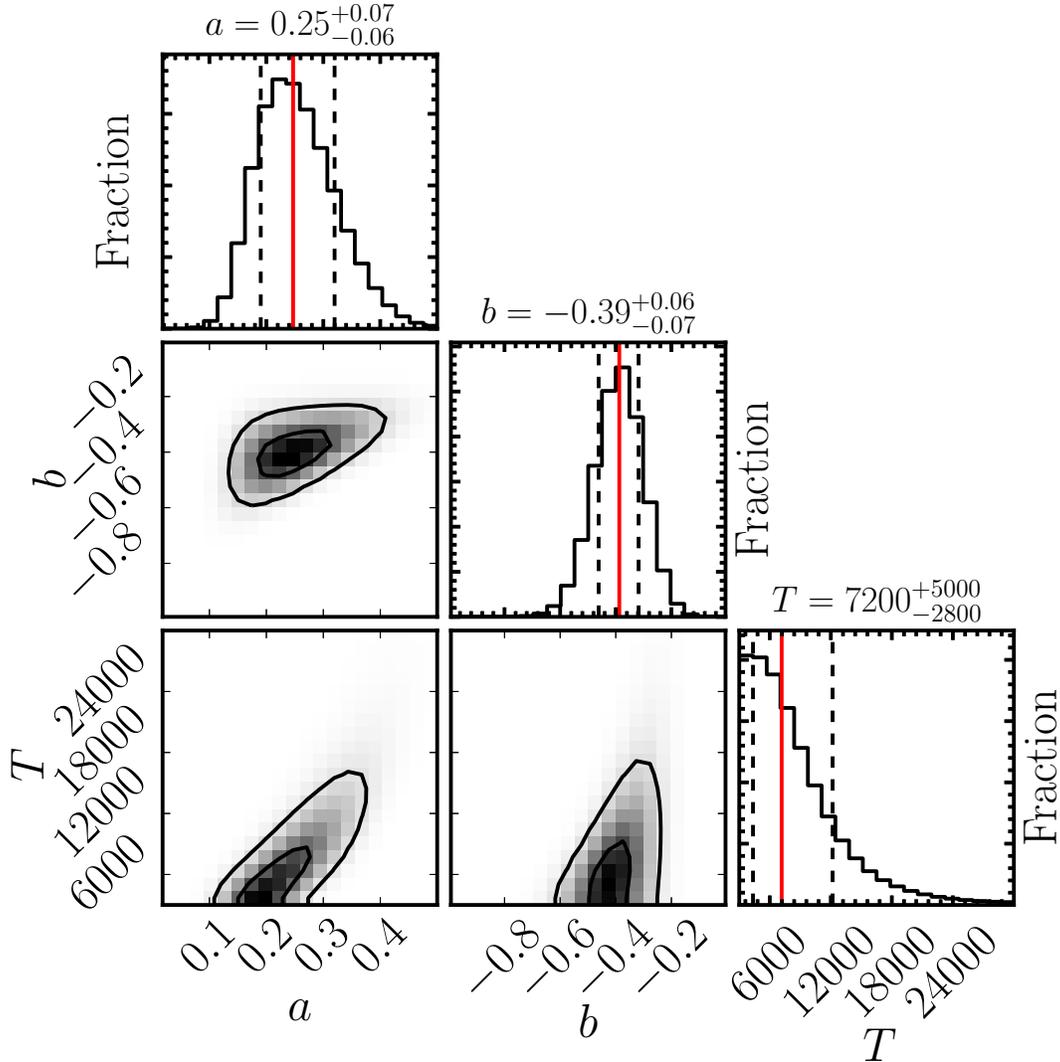}
\end{center}
\caption{Posterior probability distributions for model parameters $a$, $b$, and T, and the correlations between them. We display the median value (red vertical line),  lower  error estimated by the 16\% value and higher error estimated by the 84\% value (dashed black vertical line) for the model parameters $a$, $b$, and T. The two contours in the correlation plot represent 68\% (small one) and 95\% (big one) confidence levels. }
\label{fig:mcmc}
\end{figure*}

In Figure \ref{fig:mcmc} we present  the posterior distributions of the model parameters $\Theta = (a, b, T)$, as well as the median value and the one $\sigma$ confidence level (based on the 16th and 84th percentiles), and the correlations between them. The preferred value of $a$, 0.25$^{+0.07}_{-0.06}$,  indicates that the cool gas fraction for $\sim$ Milky Way size galaxy with halo mass $10^{12} M_\odot$ is 25$^{+7}_{-6}$\%.
The preferred value of $b$, $-0.39^{+0.06}_{-0.07}$, confirms that the cool gas fraction decreases significantly with increasing halo mass. 
The median value of temperature with one $\sigma$ confidence level is 
$7200^{+5000}_{-2800}$ K,
consistent with both our previous estimation \citep{zhang2016} and the bimodal distribution of temperature with one peak centered on $10^4$ K \citep{Haider2016,Cui2019}. {\bf This is not only the most likely fit, but is in absolute terms a good fit, with a $\chi^2_\nu = 0.87$, which can only be rejected with $<$ 0.5\% confidence.}

In the analogous model for the cool gas fraction as a function of stellar mass, $M_*$,  is scaled by $10^{10} M_\odot$ and the relation is $f_{\rm cool}(M_*) = a \times (\frac{M_*}{10^{10}M_\odot})^b$, The preferred value of $a$, 0.28$^{+0.07}_{-0.06}$,  indicates that the cool gas fraction for $\sim$ a galaxy with stellar mass $10^{10.0} M_\odot$ is 28$^{+7}_{-6}$\%. 
The preferred value of $b$, and $T$ are  $-0.33\pm0.06$, and 
%10865^{+9662}{_{5263}
$9500^{+7100}_{-3600}$ K,
respectively.

\subsection{Comparison to Previous Results}

\begin{figure}[htbp]
\begin{center}
\includegraphics[width = 0.48 \textwidth]{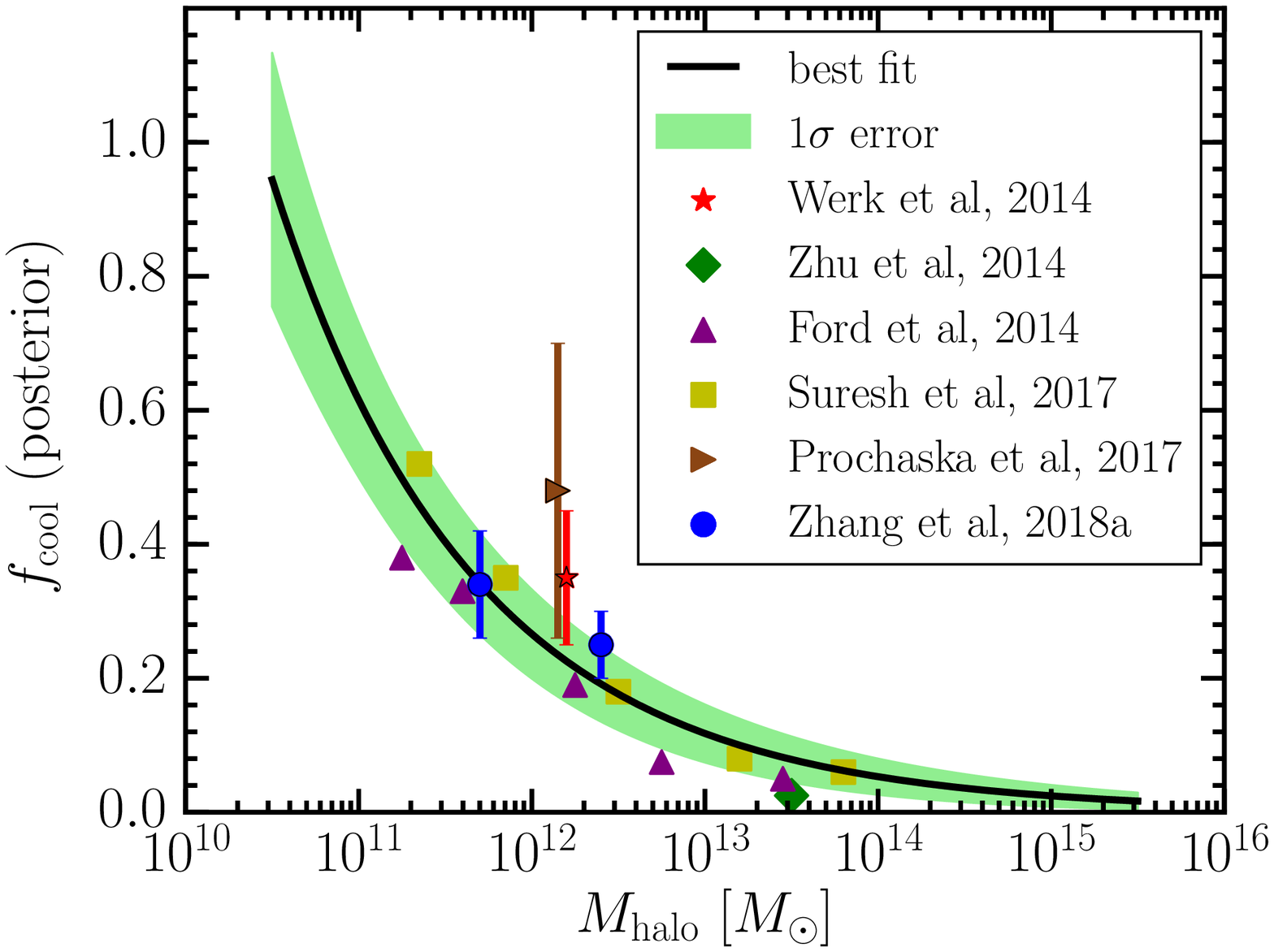}
\end{center}
\caption{The posterior cool gas fraction as a function of the halo mass. The shaded region indicates the 1$\sigma$ uncertainty estimated from the entire MCMC sample. The data points from both \cite{Ford2014} and \cite{suresh2017} are corresponding to gas with temperature $< 10^5$ K.}
\label{fig:coldgrac}
\end{figure}

We show the posterior distribution for the cool gas fraction of the CGM, within projected radii of $0.05 < r_s < 0.25$, as a function of halo mass in Figure \ref{fig:coldgrac}. If we assume that this fraction is relatively independent of the radius over which it is measured, then we can easily compare the results to those from other studies and simulations. 
We compare first to our own previous estimates (Paper II) of $0.34\pm 0.081$ for low mass halos ($M_h \approx  10^{11.7}$ M$_\odot$) and $0.26 \pm 0.05$ for high mass halos ($M_h \approx 10^{12.4}$ M$_\odot$). These values are consistent with our current results.

Next, we compare our results to estimates from various published studies in Figure \ref{fig:coldgrac}. We plot cool gas fraction measurements derived from absorption line observations by \cite{Werk2014}. They present a range of $f_{\rm cool}$ for gas at $T\sim 10^4$K from 0.25 to 0.45. We adopt $f_{\rm cool} = 0.35\pm0.1$ for halos of mass $10^{12.2} {\rm M}_\odot$ to represent their result, {\bf which is slightly more than one $\sigma$ away from our prediction ($0.21\pm0.06$).}  
\cite{prochaska2017} report a higher value of $f_{\rm cool}$ based on absorption line measurements for similar mass galaxies ($M_h \sim 10^{12.15} M_\odot$) 
%$(9.2\pm4.3)\times 10^{10} M_\odot$, which corresponds to cool gas fraction 
of $0.48\pm0.22$, {\bf  which is also slightly more than one $\sigma$ away from our prediction ($0.22\pm 0.07 $)}.
\cite{Zhu2014} find that the cool gas traced by Mg II in the halos of luminous red galaxies with mass $\sim 10^{13.5} {\rm M}_\odot$ is between $10^{10}$ and $10^{11} {\rm M}_\odot$, assuming 0.1 solar abundance. This measurement corresponds to cool gas fractions of $0.0025$ to $0.025$. 
%{\bf Jess: Love Figure 5...I think your representation of my result is great. But I Would actually look at Prochaska et al. 2017...I think his fraction is ALOT higher (we had better constraints on HI in that paper), which might be inconsistent with your fit here... and thus interesting. The errors might be better constrained in that paper. I think you can also compare your results to the LRG (Luminous Red Galaxy) work of Zhu+ (I think 2014?), and Sean Johnson/Hsaio-Wen Chen have a result too...It's only MgII, but I think they make some quick mass estimates in there, so you can observationally constrain the high mass end. I believe this MAY be inconsistent with your best fit too!! In fact, the work with LRGS (Prochaska+student's work) finds that more massive galaxies have a HUGE amount of gas in the cool phase, and it's super weird. It's worth reading these papers for context. I think you really should expand this section to compare with observations. I'll send an email with some links to these papers. }

Given the current large uncertainties, {\bf we consider} our results {\bf to be} statistically consistent with previous observational results. {\bf Nevertheless, one might argue that there is } a slight tension at $\sim L^*$ mass scales between the absorption line results and ours {\bf given the slightly larger than 1 $\sigma$ differences}, where the absorption line measurements indicate larger $f_{\rm cool}$. This tension may reflect the difference in approaches, in which we actively account for contributions from neighboring galaxies both by limiting the projected radius and including the contribution of neighboring galaxies in our modeling.

Comparing to results from simulations is a bit more complicated. We were able to extract values for the cool gas fraction from both \cite{Ford2014} and \cite{suresh2017} but these values  correspond to all gas with temperature $<10^5$ K. As such, these values may somewhat overestimate their predicted fractions of gas at $T \sim 10^4$K. With this caveat in mind, we still find excellent agreement both in the normalization and in the behavior of $f_{\rm cool}$ with $M_h$. There are a number of other simulations of the CGM \citep[e.g.,][]{Oppenheimer2016}, but it is is often difficult to extract quantities that are directly comparable to our observations. We encourage future simulators to present predicted line emission ``images" of the CGM.

\subsection{Caveats}
\label{sec:caveats}

There are a number of complicating factors in our measurements of cool gas fraction of the CGM and the interpretation of our empirical results.

\subsubsection{The $f_{\rm cool}$ and T Model Degeneracy}

Because the emission flux depends on both the cool gas fraction $f_{\rm cool}$ and temperature $T$ (Eqs. \ref{eq:recombination} and \ref{eq:halo mass}), those two parameters are somewhat degenerate in our modeling. So far, however, we have not allowed $T$ to vary with $M_h$ so there is a concern that some of the derived variation of $f_{\rm cool}$ on $M_h$ could rather be due to a variation of $T$ with $M_h$.

One option is to allow both $f_{\rm cool}$ and $T$ to vary with $M_h$ in our modeling. Unfortunately, we have only a few data points to fit and as such the resulting constraints are poor. Nevertheless, we explore 
the alternative modeling scenario  where both $f_{\rm cool}$ and $T$ have a power law dependence on $M_h$. The resulting fit has $f_{\rm cool}$ with $a = 0.25\pm0.08$ and $b=-0.37^{+0.06}_{-0.07}$, which  is almost indistinguishable from the previous estimation, while {\bf for the associated power law model for} $T$ {\bf it results in} $a = 1.63^{+2.72}_{-1.10} \times 10^4$ {\bf and} $b=1.52^{+0.36}_{-0.87}$. $T$ %dependence 
is poorly constrained.% (no variation is within the 2$\sigma$ confidence limit). 
We conclude that our initial adoption of a single value for $T$ is not significantly affecting the behavior we are inferring for $f_{\rm cool}$ as a function of $M_h$.

%\subsubsection{Scatter in Stellar Mass - Halo Mass Relation}

%{\bf
%In our analysis, we have used the \cite{yang2007, Yang2012} estimates of the halo masses based on the rank of the system's total stellar mass.
%As demonstrated by \cite{yang2007} and \cite{Lim2017} using mock SDSS galaxy and group catalogs, these halo mass estimates have a scatter that ranges between 0.2 to 0.3 dex, depending on the galaxy formation model one adopts. We have not yet considered the effect of this scatter. Here, we adopt 0.3 dex Gaussian scatter on the halo mass estimation to assess the impact on our results. We ran 1000 iterations incorporating this halo mass scatter, in which we measured the flux in each mass bin. The median flux for each of the four mass bins across the 1000 simulations is 0.013, 0.0021, 0.0011, and 0.00038 for the inner radius, and 0.0027, 0.0018, 0.00075, and 0.00018 for the outer radius. %These  fluxes measured when scatter on the halo mass added are consistent with the results in Table \ref{tab:Rdata}. 
%When we reassess our model parameters in Eq. \ref{eq:halo mass}, we find:
%$a = 0.18^{+0.08}_{-0.05}$ and $b = -0.48^{+0.12}_{-0.13}$, consistent with the previous values of
%$0.26^{+0.07}_{-0.06}$ and $-0.37^{+0.06}_{-0.07}$, respectively.
%In general, the scatter for the halo mass is thought to be smaller than 0.3 dex, especially at the low and high mass ends of the range, so we expect that the actual effect is smaller than which we present here.}

\subsubsection{Clumpiness}

We assume a smooth gas distribution characterized by an equilibrium gas distribution within an NFW potential. The adoption of both the equilibrium model and the smooth density distribution are questionable. Further guidance on these topics from detailed simulations is an urgent need. However,
the agreement in the overall observed radial emission profile \citep{zhang2018a} suggests that the density profile is not likely to be grossly different from what we have adopted. Even so, local
changes in gas density are  problematic as the emission depends on density squared and can affect the overall normalization of the profile.

In our models, we only change the fraction of gas in the cool component, $f_{\rm cool}$. However, an additional degree of freedom that should be considered is the clumpiness of this cool gas. We have discussed this degree of freedom previously \citep{zhang2018a}. While we have no additional insights into modeling this behavior, we note that on the basis of our comparison to absorption line results and simulations (Figure \ref{fig:coldgrac}) we would not expect clumpiness to significantly alter our derived values of $f_{\rm cool}$. Clumpiness would in general lower the required values of $f_{\rm cool}$ and as such might result in somewhat better agreement with the \cite{Ford2014} results, although both the observational results of \cite{Werk2014} and the simulations of \cite{suresh2017} suggest that $f_{\rm cool}$ is not much lower than what we have found in our reference model.

The degree of clumpiness could vary as a function of halo mass \citep[see][for an example of varying CGM morphology that depends on mass]{fielding},
which would affect the distribution we show in Figure \ref{fig:coldgrac}.
Clumping might, in fact, help bring the $f_{\rm cool}$ values down for the lowest mass systems.

\section{Summary}

We present measurements of the H$\alpha$ emission line flux as a function of halo or stellar mass, $M_h$ and $M_*$, respectively, at two different scaled radius for systems spanning $10^{10.7} < M_h < 10^{15}$ M$_\odot$ and $10^{8.7} < M_* < 10^{12.5}$. The flux at both radii drops as both $M_h$ or $M_*$ increase, indicating a strong inverse relationship between the cool gas fraction of the CGM, $f_{\rm cool}$ and both $M_h$ and $M_*$. Using a highly simplified model of cool gas in halos, we quantify the behavior of $f_{\rm cool}$ with with $M_h$ and $M_*$: 
%$C_f(M_h) = (0.23^{+0.07}_{-0.06}) \times (M_h/10^{12}M_\odot)^{(-0.41^{+0.07}_{-0.09})} + (0.028^{+0.030}_{-0.020})$ 
$f_{\rm cool}(M_h) = (0.25^{+0.07}_{-0.06}) \times (M_h/10^{12}M_\odot)^{(-0.39^{+0.06}_{-0.07})}$
and
$f_{\rm cool}(M_{*}) = (0.28^{+0.07}_{-0.06}) \times (M_{\rm *}/10^{10.0}M_\odot)^{(-0.33\pm 0.06)}$.
%$C_f(M_{*}) = (0.25\pm 0.07) \times (M_{\rm *}/10^{10.0}M_\odot)^{(-0.35\pm 0.07)} + (0.026^{+0.032}_{-0.019})$. 
Our prediction for the cool gas fraction is as high as 90\% for the smallest halos we probe,   $26^{+7}_{-6}$\% for a Milky Way like galaxy, and only a few percent for the most massive clusters.

These results are consistent with results from previous absorption line studies \citep[$0.35\pm0.1$ for $M_h \sim 10^{12.2}$ M$_\odot$;][]{Werk2014} and simulations \citep{Ford2014,suresh2017}, where comparison is possible. 

The quantitative results we present are tentative in the sense that they rely on highly simplified models. There are a number of complications that we neglected, principally the clumpiness of the gas, 
%{\bf the scatter in our estimates of the halo mass,}
 deviations in the gas distribution from the hot gas in thermodynamical equilibrium, and the assumption of a constant cool gas fraction across radii. The agreement in the derived $f_{\rm cool}$'s with absorption line studies and simulations suggests that our simplifications do not grossly affect our inferences, but our measurements warrant detailed comparison to results from more realistic CGM models. Stacking is providing a range of new constraints on the cool CGM \citep{zhang2016,zhang2018a,Zhang2018b,Zhang2019}, with the potential to provide even more both from the further analysis of SDSS spectra and from the future analysis of DESI \citep{desi} data. We strongly encourage simulators to provide predicted CGM properties projected onto this observational space and to identify what observations will best distinguish between models.

\section{Acknowledgments}

DZ and HZ acknowledge financial support from NSF grant AST-1713841. This work is supported by the 973 Program (Nos. 2015CB857002) and national science foundation of China (grant
Nos. 11833005, 11890692, 11621303). We also thank the support
of the Key Laboratory for Particle Physics, Astrophysics and
Cosmology, Ministry of Education. The authors gratefully acknowledge  the SDSS III team for providing a valuable resource to the community.
Funding for SDSS-III has been provided by the Alfred P. Sloan Foundation, the Participating I institutions, the National Science Foundation, and the U.S. Department of Energy Office of Science. The SDSS-III web site is http://www.sdss3.org/.

SDSS-III is managed by the Astrophysical Research Consortium for the Participating Institutions of the SDSS-III Collaboration including the University of Arizona, the Brazilian Participation Group, Brookhaven National Laboratory, Carnegie Mellon University, University of Florida, the French Participation Group, the German Participation Group, Harvard University, the Instituto de Astrofisica de Canarias, the Michigan State/Notre Dame/JINA Participation Group, Johns Hopkins University, Lawrence Berkeley National Laboratory, Max Planck Institute for Astrophysics, Max Planck Institute for Extraterrestrial Physics, New Mexico State University, New York University, Ohio State University, Pennsylvania State University, University of Portsmouth, Princeton University, the Spanish Participation Group, University of Tokyo, University of Utah, Vanderbilt University, University of Virginia, University of Washington, and Yale University.

\bibliography{bibliography}

\end{document}